\documentclass[preprint2]{aastex} 

\usepackage[dvips, bookmarks, colorlinks=true, pdftitle={Update on Pluto Satellites project}, pdf
author={Kevin Walsh}, pdfsubject={Collisionally damping disks around Pluto}, pdfkeywords={Pluto, 
Charon, Nix, Hydra, Symba}]{hyperref}
\usepackage{graphicx}
\usepackage{amsfonts}
\newcommand{\hide}[1]{} 

\begin{document}
\title{Formation and evolution of Pluto's small satellites}
\author{Kevin J. Walsh,  Harold F. Levison}
\affil{Southwest Research Institute, 1050 Walnut St. Suite 300, Boulder, CO 80302, USA}
\email{kwalsh@boulder.swri.edu}

\begin{abstract}

Pluto's system of 5 known satellites are in a puzzling orbital
configuration. Each of the four small satellites are on
low-eccentricity and low-inclination orbits situated near a mean
motion resonance with the largest satellite Charon.   The Pluto-Charon
binary likely formed as a result of a giant impact and so the simplest
explanation for the small satellites is that they accreted from debris
of that collision.  The Pluto-Charon binary has evolved outward since
its formation due to tidal forces, which drove them into their current
doubly synchronous state. Meanwhile, leftover debris from the
formation of Charon was not initially distant enough from Pluto-Charon
to explain the orbits of the current small satellites. The outstanding
problems of the system are the movement of debris outward and the
small satellites location near mean motion resonances with Charon.

This work explores the dynamical behavior of collisionally interacting
debris orbiting the Pluto-Charon system. While this work specifically
tests initial disk and ring configurations designed to mimic the
aftermath of the disruption of satellites by heliocentric impactors,
we generally find that collisional interactions can help move material
outwards and keep otherwise unstable material dynamically bound to the
Pluto-Charon system.  These processes can produce rings of debris
whose orbits evolve rapidly due to collisional processes, with
increasing pericenters and decreasing semimajor axes. While these
rings and disks of debris eventually build satellites significantly
further out than the initial locations of a disrupted satellite, they
do not show a strong preference for building satellites in or near
mean motion resonances with Charon under a wide array of tested
conditions.
\end{abstract}

\keywords{minor planets, asteroids: formation --- Kuiper Belt objects: general ---  planets and satellites: dynamical evolution and stability}

\section{Introduction}

Charon is the most massive satellite in the Solar System relative to
it's primary's mass, with $M_{\mathrm{Charon}} = 0.1126
M_{\mathrm{Pluto}}$ (Beauvalet et al. 2013). It is a doubly
synchronous system, with rotational and orbital periods of 6.38~days,
orbiting on a near-zero eccentricity orbit with semimajor axis
$a_\mathrm{Charon}=19596$~km, which is $\sim$17~$R_\mathrm{Pluto}$
(Stern et al. 2003, Buie et al. 2012, Brozovi\'{c} et al. 2014). Pluto
has at least four smaller satellites, each orbiting with a period that
is near an integer ratio of Charon's period and at distances between
30--60~$R_\mathrm{Pluto}$ (Weaver et al. 2006, Buie et al. 2006, Stern
et al. 2007, Showalter et al. 2011, Showalter et al. 2012, Buie et
al. 2013, Brozovi\'{c} et al. 2014). Their orbits are estimated to be
nearly circular ($e<0.01$), and co-planar ($i<1^\circ$) (Brozovi\'{c}
et al. 2014).

While there are doubly synchronous binary asteroid systems with
  even higher mass ratios (where satellite mass divided by primary
  mass is closer to 1; Main Belt Asteroid (90) Antiope and Trojan
  Asteroid (617) Patroclus are both nearly similarly sized; Richardson
  and Walsh 2006), the seemingly delicate dynamics of the system of
  small satellites are not found duplicated among asteroid satellite
  systems. Meanwhile the extreme size of Charon relative to Pluto
  makes it unique in the Solar System relative to planetary
  satellites, but the system has some similarity, at least in
  complexity, to Saturn's system. Some works, including this one,
  envision the small satellites forming from a disk or ring of
  debris. This disk would differ from Saturn's rings, or the recently
  discovered rings around the Centaur (10199) Chariklo (Braga-Ribas et
  al. 2014), as it would be entirely outside the Roche Limit of Pluto
  where Styx, the innermost small satellite, is over 10 times further
  from Pluto than its nominal Roche Limit. Therefore particles could
  accrete into large satellites once relative velocities became low on
  timescales that are fast compared to viscous stirring timescales.

\subsection{Physical and Orbital Properties}

Brozovi\'{c} et al. (2014) reported the best fits for the orbital
properties of Nix, Kerberos and Hydra, based on multiple Hubble Space
Telescope observing campaigns (Table \ref{satprops}; see also Buie et
al. 2013). Of particular importance to this study are the period
ratios between each satellite and Charon; 3.1565, 3.8913, 5.0363,
5.9810 for Kerberos, Nix, Kerberos and Hydra (Brozovi\'{c} et
al. 2014). From these data it appears that none of the satellites are
currently in resonance with Charon or each other, and they are not
systematically on the inside or outside of resonance, nor do their
distances from resonance correlate with size.

The physical properties of the small satellites are difficult to
estimate since imaging is limited to optical wavelengths and
photometric measurements can determine size only with an estimate for
each bodies' albedo, and mass only with an estimate for density and
size. Orbital stability has been used to place limits on masses of
both hypothetical satellites (pre-discovery) and also for subsets of
the current system.  Stern et al. (1994) studied the stability of
hypothetical satellites in the Pluto-Charon system, primarily placing
an upper limit on satellite mass, $3\times10^{-4}$(M$_\mathrm{Pluto}$
+ M$_\mathrm{Charon}$), that would create observable perturbations in
Charon's orbit. Following the discovery of Nix and Hydra, Pires dos
Santos (2011) tested for stable regions where more satellites could
reside assuming masses of 5.8$\times10^{17}$~kg and
3.2$\times10^{17}$~kg for Nix and Hydra respectively (using masses
from Tholen et al. 2008). With these masses the stable region between
those satellites is quite narrow and centered around the 5:1 MMR --
the location where Kerberos was eventually found.

Youdin et al. (2012) presented a series of numerical experiments
attempting to constrain the mass of the larger satellites (Nix and
Hydra) by considering the long-term survival of Kerberos, orbiting
between them. This work suggested that period ratios for Kerberos
relative to Charon were more stable below 4.98 and above 5.01, which
agrees with the recent period ratios estimate (5.0363) from
Brozovi\'{c} et al. (2014). Upper limits for mass found in that work,
$M_\mathrm{Nix}\lessapprox 5\times 10^{16}$~kg and
$M_\mathrm{Hyd}\lessapprox 9\times 10^{16}$~kg, which required that
the satellites have an albedo above 0.30 for the assumption of an
internal density of 1~g~cm$^{-3}$.

The orbital fits by Brozovi\'{c} et al. 2014 also constrained the
masses of individual satellites. The larger satellites were found to
have masses $M_\mathrm{Nix}= 4.5\times 10^{16}$~kg and $M_\mathrm{Hyd}
= 4.7\times 10^{16}$~kg, which are both similar and within a factor of
two respectively from that found in Youdin et al. (2012). The
estimated mass for Kerberos is $M_\mathrm{Kerberos}= 1.6\times
10^{16}$~kg. Using a range of possible visible albedo (35\%--4\%),
this work estimated radii ranges for all four satellites: 4--14~km,
23--70~km, 7--22~km, and 29-86~km for Styx, Nix, Kerberos and Hydra
respectively.

\begin{table}[h]
\begin{tabular}[h]{l|lllllll}
Satellite & $a$ (km)   & $a$ (R$_\mathrm{Pluto}$)     & $e$        & $i$ (deg) & P (days)    & P$_\mathrm{sat}$/P$_{\mathrm{Charon}}$   &R (km)  \\ 
\hline 
Charon    & 19596  & 16.59  & 0.00005     &0.0     & 6.3872  & 1 & 603.6$\pm$1.4 \\ 
Styx      & 42413  & 35.91  & 0.00001     &0.0     &20.1617  & 3.1565   & 4--14 \\
Nix       & 48690  & 41.22  & 0.00000     &0.0     &25.8548  & 3.8913    & 23--70 \\ 
Kerberos  & 57750  & 48.89  & 0.00000     &0.4     &32.1679  & 5.0363   & 7--22 \\ 
Hydra     & 64721  & 54.80  & 0.00554     &0.3     &38.2021  & 5.9810   & 29--86 \\
\end{tabular}
\caption{ Orbital elements for all satellites and for the radii of the
  small satellites are from Brozovi\'{c} et al. (2014). Charon's
  diameter is from occultation measurements (Sicardy et
  al. 2006). Pluto's radius, 1181~km, for $a$/R$_\mathrm{Pluto}$
  estimates from Lellouch et al. (2009). Charon's orbital elements are
  plutocentric, while the small satellites orbital elements are
  relative to the Pluto-Charon barycenter. }\label{satprops}
\end{table}

\subsection{Formation of Charon and debris}\label{CharonForm}

Charon is thought to have formed in a giant impact (McKinnon 1989,
Canup 2005,2011). SPH simulations show that it likely remained intact
following the collision, making it a prototype for an ``intact
capture'' type of high impact-parameter collision event (Canup
2005). Formation from a disk following the impact appears to still be
dynamically possible (Canup 2005), but the intact capture is preferred
due to the fact that Charon has a similar density as Pluto.  More
violent events that result in the accretion of Charon from a
circum-Pluto disk (similar to the Earth-Moon scenarios) lead to the
loss of rocky material to Pluto and create larger density disparity
between Pluto and Charon (see Desch 2014 for a model to preserve
density similarities during the formation of Charon from a disk).

The high angular momentum of the system, and its doubly synchronous
state, point to past tidal evolution from a closer orbit of Charon
around a more rapidly rotating Pluto (Farinella et al. 1979, McKinnon
1989, Dobrovolkis et al. 1989, Cheng 2011).  The SPH models of
the Charon-forming event that satisfy these angular momentum
constraints typically find that its orbit would initially have been
eccentric (Canup 2005, 2011).  The possible post-collision orbits for
Charon are discussed in detail in Canup (2005,2011), where the initial
orbits range between semimajor axes of 4-10~$R_{\mathrm{Pluto}}$, with
eccentricities between 0.1--0.8 (see Fig 5. Canup 2011). There are
also simulations that produce $e\sim$0.9 with
$a/R_{\mathrm{Pluto}}\sim$25.

 Peale et al. (2011), Cheng (2011) and Cheng et al. (2014a), show that
 the tidal evolution of Charon required a few million years to reach
 its current semimajor axis with variations in the orbit evolution
 dominated by the choice of tidal models.  These recent works consider
 an initially eccentric orbit for Charon as predicted in the ``intact
 capture'' formation models and they find similar evolution timescales
 as previous estimates for Charon evolving on a circular orbit
 (Farinella et al. 1979, Dobrovolskis et al 1989, Peale
 1999). Solutions where the eccentricity of Charon remained moderate
 were desireable outcomes to test the viability of resonant transport
 of the satellites in the co-rotation resonance with Charon (Ward and
 Canup 2006, Lithwick and Wu 2008, Cheng 2011, Cheng et al. 2014a,b).
 Cheng et al. 2014a find for Charon to remain moderately eccentric
 during its tidal evolution the ratio of dissipation between Pluto and
 Charon has different values depending on the tidal model
 used. However, for a range of these parameters, both tidal models
 tested by Cheng et al. 2014a find evolutions where Charon has
 $e\sim$0.1--0.5 for the entire outward evolution, only damping to 0
 when Charon has reached its current semimajor axis.

There are alternative orbital evolutions due to different initial
conditions or tidal parameters. For example, Canup (2011) finds some
collision simulation outcomes where Charon has a semimajor axis close
to where it is found today, but with a much larger eccentricity
  (see Figure 5 of Canup 2011). Similarly, if the tidal dissipation
parameters were substantially different than those used in the
calculations the timescale for the evolution could change. While these
extremes are not essential for the work presented here and therefore
not discussed in depth, it is possible that other mechanisms for
explaining the small satellites may require them.

Canup (2011) characterized the debris created in the ``intact
capture'' models in terms of the mass and maximum equivalent circular
orbit of disk material ($a_{\mathrm{eq,max}}$), where $a_\mathrm{eq}$
is the circular orbit containing the same amount of angular momentum,
and $a_{\mathrm{eq,max}}$ is the maximum value for all debris in the
simulation (see Fig 6.).  The total mass of debris was typically
between $10^{17}$--$10^{21}$~kg on orbits between
$a_{\mathrm{eq}}$=2--30~$R_{\mathrm{Pluto}}$, with no clear correlation between the
two. Typical orbits of post-impact Charon are eccentric with
$a\sim$4--10~$R_{\mathrm{Pluto}}$. Holman and Weigert (1999) studied
stability around binary systems and find that in a best-case scenario
of zero eccentricity for both Charon and the debris the closest stable
orbit is at $\sim$1.974~$a_\mathrm{Charon}$. For a nominal formation
outcome of $a_\mathrm{Charon} \sim 4-10$ $R_{\mathrm{Pluto}}$, debris
closer than $a_\mathrm{Charon} \sim 8-20$ $R_{\mathrm{Pluto}}$, would be
immediately unstable. If Charon had an initially eccentric orbit then
the nearest stable orbit is further away with less debris being
stable.

The current orbits of the small satellites
($\sim$30--60~$R_{\mathrm{Pluto}}$), are much further from Pluto than
any of the debris in the simulations of Canup et al. (2011), and thus
it is unlikely that today's satellites formed directly on today's
orbits during the formation of Charon. Any satellites formed during
the formation of Charon would later witness its outward tidal
evolution. As Charon moved outward the locations of its mean motion
resonances would also move and would sweep over the orbits of any
existing satellites. Mean motion resonances with Charon would perturb
or capture satellites resulting in rapid and destabilizing
eccentricity excitement (Ward and Canup 2006).  Since tidal forces are
far too weak to damp the current satellites eccentricities (Stern et
al. 2006), their near zero eccentricity orbits also suggest that they
did not witness the tidal evolution of Charon in their present
configuration.

A corotation resonance capture has been proposed as a means to
transport satellites without exciting their eccentricities (Ward and
Canup 2006). This resonance capture requires a moderate eccentricity
for Charon during its outward migration. However, the restrictions on
the corotation resonance is that transport in resonance requires a
narrow range of eccentricity for Charon for each different resonance,
and for multiple satellites these requirements are probably mutually
exclusive (Lithwick and Wu 2008, Cheng 2011, Cheng et
al. 2014b). With a moderate or large eccentricity for Charon it is
also possible to capture a satellite in multiple resonances at once
and keep a relatively low eccentricity for the satellite. However,
this was found to be an extremely low probability event and also
ineffective for the inner resonances where Styx and Nix are found
(Cheng 2011, Cheng et al. 2014b).

Kenyon and Bromeley (2014) estimate that primordial debris from the
Pluto-Charon forming collision would, by way of a collisional cascade,
reach stable orbits just outside the region of orbital instability
caused by Charon's orbit. This ring would then spread on timescales of
5--10 years due to the tidal input of angular momentum into an
optically thick disk. These short timescales, if less than accretion
timescales, could permit spreading of the ring into a disk before
accretion begins. An essential aspect of the analytical estimate of
the spreading time is that angular momentum transfer among ring
particles slows their precession, which helps to maintain low-velocity
collisions. This avoids collisional fragmentation on timescales short
compared to that for spreading. However, Kenyon and Bromeley (2014) do
not account for the fact that low-velocity collisions will simply
result in accretion and growth.  Rather, accretion will likely happen
before spreading and a disk will not form, something that is found in
simulations presented below in Section \ref{s:lipad}. The timescales
for this disk spreading and satellite growth proposed by Kenyon and
Bromeley (2014) are faster than the tidal evolution of Charon. Any
satellite system formed in this manner, immediately following the
formation of Charon, would then be subjected to the sweeping resonances
caused by Charon's outward tidal evolution that would induce
significant eccentricities into the small satellites.

Material can also be captured from heliocentric orbit. Two passing
objects colliding within the Hill Sphere of the Pluto-Charon system
can result in capture of some of the collisional debris (Pires dos
Santos et al. 2012). The total amount of captured material depends on
the orbital and size distributions of the passing material. Pires dos
Santos et al. (2012) explored this mechanism finding that large
objects that would carry significant mass have collision timescales
that are too long, resulting in a very small total amount of captured
mass.

How the present satellites or their building blocks got to where they
are today is still an open question. The capture of material from
outside the Pluto-Charon system is inefficient, and moving the present
day satellites in various arrangements of orbital resonances has not
been demonstrated.  Moving a single satellite outward may be possible
(Lithwick and Wu 2008; see also Cheng 2011, Cheng et al. 2014b), and
some eccentricity excitement would be irrelevant if the transported
satellite simply served as the source material for the entire suite of
today's satellites (of course too much eccentricity excitement can
lead to dynamical ejection or accretion by Pluto or Charon). A
satellite disrupted after the tidal evolution of Charon could form a
collisionally active disk as envisioned by Kenyon and Bromeley
(2014). But growth of satellites from a collisionally damped disk does
not imply growth near mean motion resonances, as Kenyon and Bromeley
(2014) found no strong preference for growth in those locations.

The breakup of a satellite  may also form an eccentric ring,
which could have strong dynamical interactions with Charon at
the same time as it is collisionally evolving. Similarly, the
disruption of a primordial satellite {\it during} the tidal evolution
of Charon could have interactions with Charon while it is still on an
eccentric orbit.  If a satellite is disrupted it will be important
  if the dynamical environment forces it to re-accrete in a new
  location.  If the timescale for the disruption and reaccretion of a
  satellite is short compared to the tidal evolution of Charon, then
  it is a process that could be repeated multiple times.  This could
  be a way to avoid dynamical ejection of a satellite as repeated
  disruption and re-accretion events could help it avoid interacting
  with the strongest resonances that sweep outward as Charon's orbit
  expands.  What happens during its disruption, its evolution as a
  ring or disk of debris and its reaccumulation into a new satellite
  could then be important.  Here we endeavor to study the evolution
of debris following the disruption of a satellite.

\subsection{The role of collisional evolution}

The state of the Kuiper Belt at the time of the Charon-forming
collision is important for determining the collisional environment of
the Pluto-Charon system during and after its formation. The giant
impact that formed Charon was the last giant impact on Pluto, but must
characterize a substantially different collisional environment in the
Kuiper Belt than is found today. Today the chance of such a collision
is essentially zero (Brown et al. 2007). Meanwhile the tidal evolution of
Charon was relatively short compared to timescales for dynamical
mixing and depletion of the Kuiper Belt (Levison et al. 2008), and
thus an enhanced collisional environment may have persisted throughout
the entire tidal evolution. In this study we will repeatedly refer to
this idea and consider the possibility that small satellites
($D<100$~km) in the Pluto-Charon system may have had a very short
collisional disruption timescales in the epoch immediately following
the Charon-forming event and {\it during} the tidal evolution of
Charon. 

Collisional interactions between particles in the system will change
their dynamical evolution by damping of excess energy and changing
particles orbits. A satellite experiencing isotropic collisions in
orbit will experience a damping of its radial velocity, decreasing
eccentricities. Meanwhile angular momentum of the system is conserved
among a collisionally active swarm, so while orbits will evolve to
lower eccentricity, $e$, the value of $a\times\sqrt{1-e^2}$ is
conserved for the population of particles resulting in decreased
semimajor axis $a$, and increased pericenter $q=a(1-e)$. Orbits with
lower $e$ and larger $q$ are more stable and longer-lived in the
Pluto-Charon system, and so even small amounts of collisional
evolution may be important for increasing lifetimes of satellites or
debris.

In this work we focus on the role that collisional evolution could
play in both the transport of material outward during the tidal
evolution of Charon, and also during the accretion of satellites
following the conclusion of tidal evolution. Unlike previous
  works we examine the outcome of satellite disruption around a
  tidally evolving Charon and the formation and evolution of the
  eccentric ring of debris. For collisional processes to be
important there must have been collisions, and material or debris in
enough quantity to affect the evolution of the system.  Specifically
we assume, and then explore the idea, that the current satellites we
see today are built from the pieces of previous disrupted satellites.

In Section \ref{s:stab} we explore stable orbits in $a-e$~space around
Pluto-Charon systems with different eccentricity for Charon's orbit
since tidal evolution models allow for a range of
eccentricities. \hide{ In Section \ref{s:cap} we explore capture
  probabilities in different MMR with Charon as a function of particle
  migration timescales and the eccentricity of both Charon and on
  migrating test particles.}  In Section \ref{s:coldisk} we explore
the collisional evolution of a few simple idealized disks of
particles.  Finally, in Section \ref{s:lipad} we model the evolution
of the debris of a disrupted satellite, and compare this
evolution among systems with different eccentricities for Charon's
orbit, and also around a system with a single central body.

\section{Stability of orbits in the Pluto/Charon system}\label{s:stab}

The long term stability of a particle around the Pluto-Charon pair
depends on the particle semimajor axis and the eccentricity of both
it's orbit and that of Charon.  The innermost stable orbit, often
cited as a critical semimajor axis ($a_\mathrm{crit}$), is a value
that has been explored in detail in previous numerical and analytical
work. Holman and Weigert (1999) produced an empirical fit for the
$a_\mathrm{crit}$ as a function of orbital eccentricity ($e$) and
reduced mass ($\mu=M_\mathrm{2}/(M_\mathrm{1} + M_\mathrm{2}))$.
Using the Pluto-Charon reduced mass $\mu=0.104$ this simplifies to the
following formulation, $a_\mathrm{crit} = 1.974 +4.65e -2.17e^2$.

Of particular interest for the origin of the small satellites of Pluto
are regions of stability while Charon is tidally evolving, during
which Charon could have a range of non-zero eccentricities. We have explored
stability limits for the Pluto-Charon mass ratios, seeking not just
the innermost stable orbit, but the envelope of allowable
eccentricities as a function of distance from Pluto-Charon for a range
of Charon eccentricity.

\begin{figure}[h]
\includegraphics[angle=0,width=3.2in]{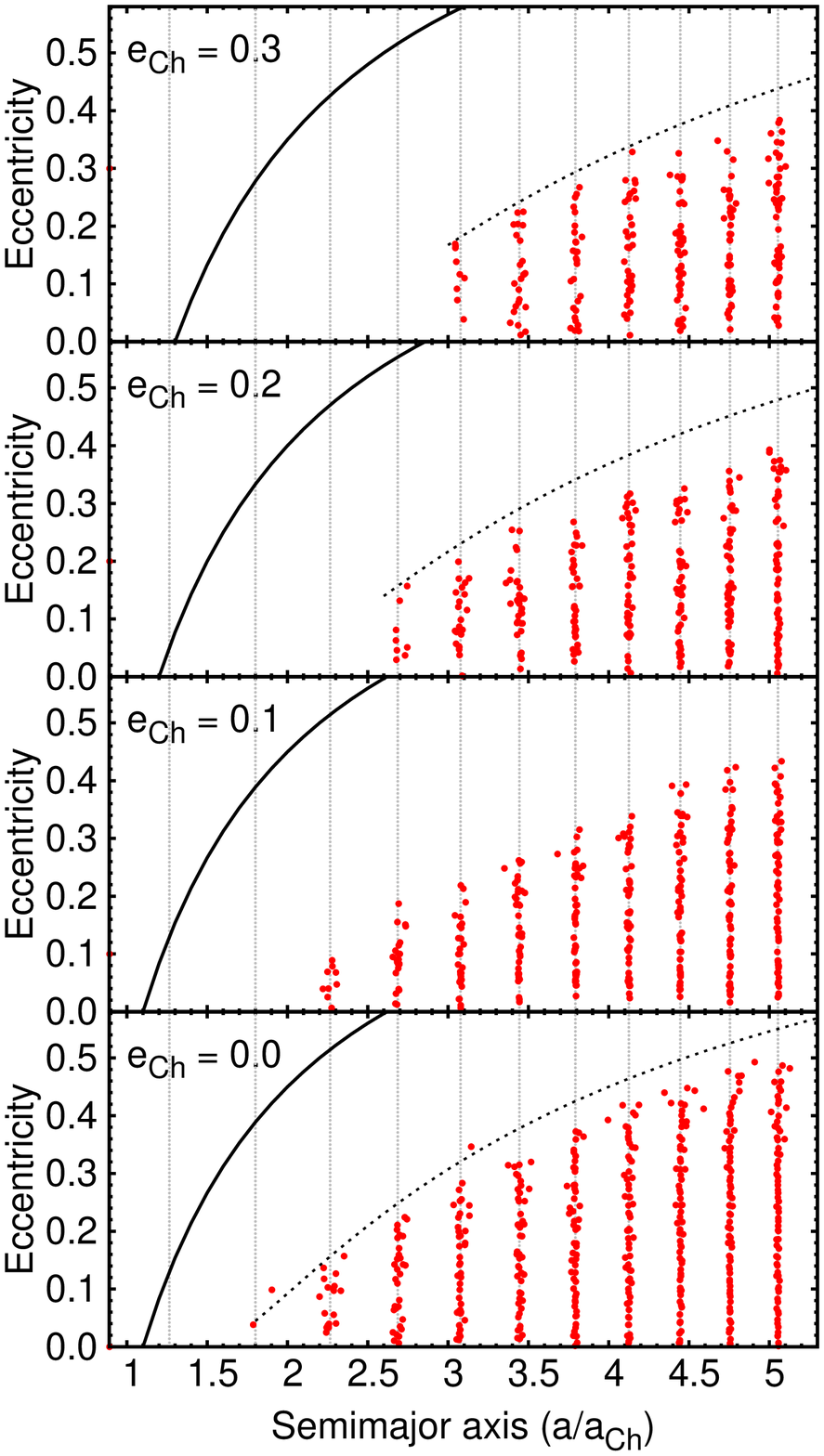}
\caption{Survival of particles orbiting Pluto and Charon as a function
  of their initial $a$ and $e$, and the eccentricity of Charon,
  $e_{\mathrm{Ch}}$. The smallest dots (black) represent the initial
  distribution of orbits for test particles at time=0 and the largest
  dots (red) show their orbits after 1000 years. Empirical curves are
  drawn as an envelope to the stable regions for a circular orbit of
  Charon (bottom frame), and for a Charon eccentricity of
  $e_\mathrm{Charon}=0.3$ (top frame). }
\label{fig:Stab}
\end{figure}

We include tests for Charon eccentricity between 0.0--0.3, in steps of
0.1 and consider initial test particle eccentricities ranging from
0--0.7, and initial semimajor axes that ranged from 1.35--5.1
$a_\mathrm{Charon}$ with randomized orbital angles (see Fig
\ref{fig:Stab}). The tests were simulated for 1000~years using the
    {\tt swift\_symba5} numerical integrator (Duncan et al. 1998) with
    a timestep of 2$\times10^{-4}$~years (1.75~hours, which is
    $\sim$1/87th Charon's orbital period). We consider co-planar or
    nearly co-planar cases as the observed system of small satellites
    all appear to be very close to co-planar (see also Pires dos
    Santos et al. 2011, who considered some small inclination
    variations in similar tests).

For each of the tested eccentricity values of Charon, $e=0.0-0.3$, the
formulation of Holman and Weigert (1999) yields
$a_\mathrm{crit}=$1.97, 2.41, 2.81, 3.17~$a_\mathrm{Charon}$.  These
values were found to agree with those found by Dvorak et
al. (1989). The simulations performed here find similar innermost
stable orbits.  For Charon on a circular orbit we find virtually no
stable orbits inside of the 3:1~MMR ($\sim$2.08~$a_\mathrm{Charon}$;
see Fig \ref{fig:Stab}), near 1.97~$a_\mathrm{Charon}$ as found in the
Holman and Weigert work. Meanwhile, at the location of the furthest
known satellite, Hydra, near the 6:1~MMR, particles are unstable with
$e>0.35$.

We have made an empirical fit to the stable region when
$e_\mathrm{Charon}=0$ with a curve described by $e < (1 - (1.7
a_{\mathrm{Charon}})/a)^{(5/3)}$. We use this curve throughout the
later sections as a guide for when regions of an eccentric disk are on
unstable orbits.
 
Cases where Charon has an eccentriciy of 0.3, the stability region is
truncated near the 5:1~MMR (at 2.9~$a_\mathrm{Charon}$, similar to the
Holman and Weigert 1999 value of 3.17~$a_\mathrm{Charon}$). Lower
eccentricities are generally required for stability, with a particle
near the 6:1~MMR requiring $e<0.25$ for this case. We estimate the
boundary of this stable region with the following emprical curve, $e <
(1 - (2.2 a_{\mathrm{Charon}})/a)^{(5/3)}
~~\mathrm{and}~~a>2.9~a_\mathrm{Charon}.$

[EDITOR: Place Fig \ref{fig:Stab} here]

An important aspect of this result is that particles with an
eccentricity above the established $a$-$e$ curve are unstable on short
timescales. However, if their collisional timescale is shorter, then a
series of collisions need only to decrease their orbital eccentricity
below these stability limits to keep them in the system for long
times.  Another implication of this experiment is that while Charon
has a high eccentricity, at times as high as 0.3 as suggested by Peale
(2011), the current orbits of Nix, P4 and P5 would not be
stable. Thus, we could expect that these bodies formed after the
eccentricity damping of Charon or were in a stable resonant state that
increased their stability.  We did not find such stable resonant
states in this work, but did not design experiments specifically for
that purpose. Cheng (2011) and Cheng et al. (2014b) explored the behavior
of particles in multiple resonances and struggled particularly to find
stable cases in the 4:1 resonance.

\section{Evolution of Collisionally evolving disks}\label{s:coldisk}

In this section we model a series of eccentric rings of debris and we
track their different dynamical evolution as a function of different
collisional environments. The role of Charon in changing the evolution
of the disk is explored by alternatively placing some disks around
single bodies with the combined mass of Pluto and Charon.  We model
the collisional interactions of the ring particles and their effects
on the ring's eccentricity and semimajor axis.

Required for these tests are the calculations of inter-particle
collision outcomes. The frequency and outcome of collisions are
strongly dependent on relative particle sizes, total system mass and
orbital distributions, all of which can change rapidly owing to
accretion, fragmentation and gravitational interactions with Charon.
Statistical calculations must be made in order to model high
collisional frequencies possible for populations of small
particles. If each particle were included directly in a simulation the
total number of particles ($N$) of any calculation would be
overwhelmingly large.  However, the dynamics of capture in MMR and
interactions with Charon demand that the simulation also accurately
model the gravitational dynamics, where timescales are controlled by
the orbital period of Charon.

The primary code we employ is {\tt LIPAD}, which stands for Lagrangian
Integrator for Planetary Accretion and Dynamics (Levison et
al. 2012). It is based on the efficient integration techniques known
as the Wisdom-Holman Mapping (WHM; Wisdom \& Holman 1991), and
specifically {\tt SyMBA} that has the added property of treating close
encounters between bodies (Duncan et al. 1998). 

 In order to represent the extremely large number of particles
 required {\tt LIPAD} relies on ``tracer'' particles.  Each of the
 tracer particles represents a large number of comparably-sized
 particles on very similar orbits. Each tracer is defined by three
 quantities, the physical radius $s$, the bulk density $\rho$ and the
 total mass of particles that the tracer represents
 $m_\mathrm{tr}$. Throughout the simulation $m_\mathrm{tr}$ and $\rho$
 do not change. When fragmentation and accretion are included the
 radius $s$ can change and thus the number of particles that the
 tracer represents will change as a function of $s$, with
 $N_\mathrm{tr}=m_\mathrm{tr}/(4/3)\pi\rho s^3$.  For all of the work
 presented here, the density of each tracer is set at 1~$g$~$cm^{-3}$.

 Collisional probabilites for tracers are calculated for each tracer
 as a function of its size $s$ and the total mass, sizes and orbits of
 its neighbors using particle-in-a-box algorithms. The outcome of the
 collision then affects the tracer particle itself, as though it were
 a planetesimal of radius $s$, so that each tracer is tracing the
 behavior of the system (see Section 2.1.1 of Levison et al. 2012 for
 more details on the tracer-tracer interactions).  Meanwhile the
 dynamics of each tracer is modeled with gravity calculations and
 other effects that are handled statistically (dynamical friction,
 viscous stirring), some of which depend on the particle's radius $s$
 and the masses, sizes and orbits of its neighbors.

A particular advantage for this problem is that the Lagrangian nature
of the code enables it to model eccentric rings (see Levison and
Morbidelli 2007, Levison et al. 2012). When a tracer is determined to
have collided with another particle, it is necessary to determine the
properties of the impactor.  The orbit that the impactor would have
had prior to impact is determined by tracking particles that have most
recently inhabited the correct regime of semimajor axis $a$ and radius
$s$. From this list of possible impactors the one with the closest
true anomaly is selected, which was shown by Levison and Morbidelli
(2007) to be critical to support asymmetries and eccentric rings.

When collisions occur {\tt LIPAD} uses a fragmentation law based on
Benz \& Asphaug (1999) to determine the outcomes. This fragmentation
law determines the expected size distribution of fragments, but the
entire distribution is not assigned to either tracer - rather a radius
$s$ is chosen for each from the distribution.  The system's size
distribution is built from a number of tracer particles with different
sizes $s$, and matches standard collisional evolution codes owing to
the statistical nature of the radius selection from a large number of
collisions (see Levison et al. 2012).

For the cases presented in this Section the collisional fragmentation
and growth of particles is not used, rather we simply explore the role
of collisional damping.

\begin{figure}[h]
\includegraphics[scale=0.35]{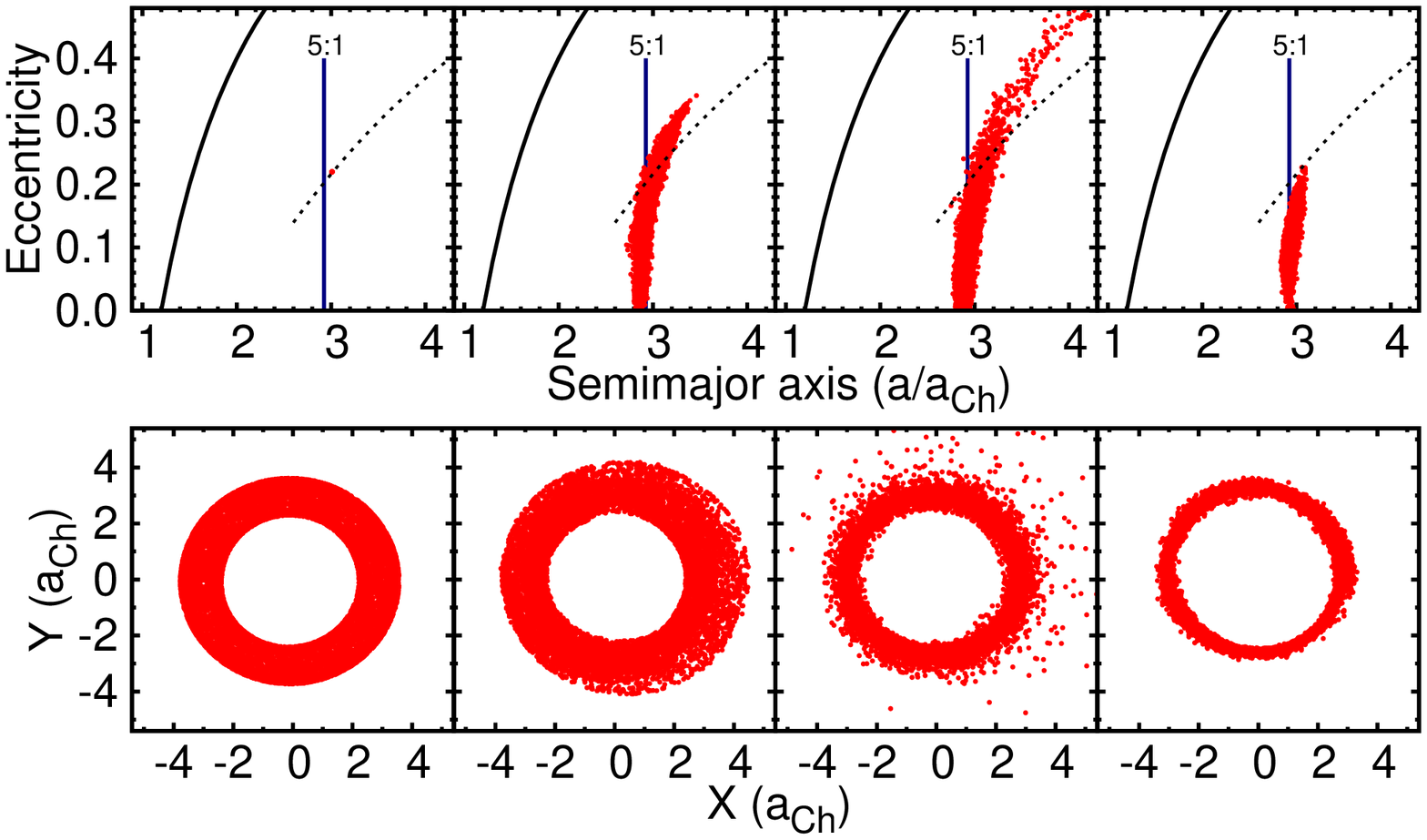} 
\caption[]{Frames showing the collisional damping of a ``simple
  eccentric disk'' of debris at times (from left to right) 0, 0.1,
  0.5, and 3.0~yr. The angular momentum of the system increases due to
  kicks from Charon and also the resonant interactions of the ring
  with Charon's orbit while the disk damps into a ring.  Note that
    the structure seen in the ring in the last frame shows that it is
    interacting with the resonance.
\label{5to1}}
\end{figure}

The first test case, dubbed ``simple eccentric disk'', is designed to
examine how the perturbations of Charon change the evolution of a
collisionally active disk at distances similar to the current small
satellites.  This test begins with a co-planar disk of particles with
the same semimajor axis $a=3.05~a_\mathrm{Charon}$ and eccentricity
$e=0.2$, and randomized orbital angles (see Fig \ref{5to1}a). The
semimajor axis is just beyond the 5:1~MMR, which is located at
$\sim$2.9~$a_\mathrm{Charon}$. The equivalent circular orbit of disk
material ($a_{\mathrm{eq}}$) is inside the 5:1~MMR at
$\sim$2.85~$a_\mathrm{Charon}$, meaning that removal of orbital energy
by collisional damping, while conserving angular momentum would result
in a ring at this lower semimajor axis orbit. All of the particles are
initially within the stability boundaries defined above in Section
\ref{fig:Stab}, and are therefore stable for long timescales. The
simulations used 10,000 particles, where each has a mass of
1.02$\times10^{14}$ kg (totalling $\sim
M_\mathrm{Nix}+M_\mathrm{Hydra}$ from Buie et al. 2008). Three
simulations were run, where the  tracer particles' representative
radii were varied to be $R$=0.3, 0.1 or 0.01~km (meaning
  $N_\mathrm{tr}= 8\times10^8, 2.4\times10^9, 2.4\times10^{10}$
  respectively). Charon's eccentricity was $\sim$0.2 throughout and
it was at its current semimajor axis.

For the case of $R=0.1$~km (illustrated in Fig \ref {5to1}) the ring
experiences rapid gravitational evolution due to interactions with
Charon. The eccentricity distribution is spread out with values
reaching as high as 0.4 in just a few Charon orbits. Collisional
damping is also dramatic for these conditions and are on similar
timescales. After a few Charon periods many particles are experiencing
damping with their eccentricities decreasing to near 0.  For these
conditions the damping effects are powerful enough to decrease all
particle eccentricities on year timescales. After this time nearly all
particles have eccentricities below their initial value of 0.2 and the
disk reaches a coherent ring-like structure (see Fig \ref
{5to1}d). 

 The evolution to a ring, an eccentric ring at first, is important for
 the results of this study. The ring does not immediately spread into
 a disk because each of the particles is very small and gravitational
 interactions can only supply extremely small changes to
 eccentricities (collisional damping timescales are much shorter
   than viscous stirring timescales). Collisions between particles
 will damp energy, and conserve angular momentum, resulting in
 decreased eccentricities.  Later, in Section 4, when accretion is
 included the ring-like structures start to spread when particles have
 grown to km-sized bodies and larger.  Note that the eccentricity of
 the particles do not damp to zero because of the presence of the
 resonance. Also, due to the collisions, no particles are lost from
 the system despite the fact that many found themselves above the
 stability curve at times.

[EDITOR: Place Figure \ref{5to1} here]

\begin{figure}[h]
\includegraphics[scale=0.35]{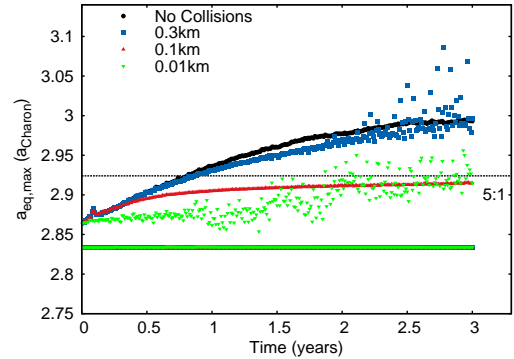} 
\caption[]{The temporal evolution of the angular momentum of our
  eccentric rings as represented by $a_{\mathrm{eq}}$. The colors
  indicate particle sizes, as shown in the legend. However, the black
  symbols show a control run where collisions were ignored. The figure
  shows two calculation for each particle size. The upper curves are
  systems with Charon. The lower curves (which are horizontal lines
  indicating that angular momentum is conserved) has a single central
  body. Note that all four of the latter runs overlap in the
  figure. The dotted line shows the location of the 5:1 MMR with
  Charon.
\label{5to1SB}}
\end{figure}

The evolution of each disk's angular momentum is correlated with the
collisional damping in each.  Initially the averaged $a_{\mathrm{eq}}$
is at $\sim$2.87~$a_\mathrm{Charon}$, just inside the 5:1~MMR with
Charon at $\sim 2.92$~$a_\mathrm{Charon}$. This value rapidly
increases, reaching $\sim 2.91$~$a_\mathrm{Charon}$ on year-long
timescales for the $R=0.1$~km case plotted in Figure \ref{5to1} (red
triangles in Fig. \ref{5to1SB}). The amount of the increase depends
strongly on the collisional evolution, which varies depending on the
representative particle radii (ranging from 0.3 km to 0.01 km).  We
also ran a simple test case without collisions, which provides an
upper bound to the outcomes shown here. The case with no collisions
experienced the most dramatic outward movement of the ring's angular
momentum owing to absence of collisions to damp particles
eccentricities (``No collisions'' in Fig \ref{5to1SB}).  There is a
clear trend in outcomes as a function of the particle radii, as the
systems with fewer collisions have the most extreme outward evolution
of system angular momentum. The system with particles $R=0.3$~km
experienced on average 0.17 collisions per particle per orbit of
Charon, which was roughly an order of magnitude less than the for the
$R=0.1$~km system, with 1.7, and roughly 3 orders of magnitude less than
the $R=0.01$~km system which had 218 collisions per particle per
Charon orbit.

[EDITOR: Place Fig \ref{5to1SB}]

A separate test removed the perturbations of Charon by examining the
same disks of debris orbiting around a single central mass of mass
equal to the combined mass of Pluto and Charon (see straight lines at
$\sim2.84$~$a_\mathrm{Charon}$ in Fig \ref{5to1SB}).  The results are
plotted together with the those described above, and are distinct as
they all show unchanging angular momentum as a function of time for
each of the same three tested radii.  This validates that, in the
  absence of external perturbation, the code conserves angular
  momentum as a population collisionally damps. It also shows that
  Charon is the cause of the angular momentum increase in our
  simulations.  

We interpret that the dominating physical effect is that the
  timescale for inter-particle collisions is on order or shorter than
  the timescale for dynamical loss. In the collisionless system we
  find that more than half of the particles are ejected from the
  system due to Charon. As the collision rates increase from zero for
  this case to a few tenths (0.17 per particle per orbit for $R
  =$~0.3~km) and up to hundreds (for $R =$~0.01~km) not only are all
  particles kept in the system, but the outward evolution of the ring
  decreases. The longer that ring particles stay highly eccentric, the
  more close interactions they have with Charon, and the more chances
  to receive kicks to expand their orbits. Thus the lower collision
  rates are able to keep particles in the system, but they are still
  kicked substantially by Charon and the ring expands. By moving to
  higher collision rates (decreasing $R$) the ring damps faster,
  allowing less time for kicks from Charon and minimal expansion of
  the rings. We will find in more complex simulations in Section 4.,
  that this outward movement of the ring works for eccentricities of
  Charon ranging from 0 to 0.3 and for a wide range of distances from
  Charon in terms if Charon's orbital separation from Pluto.

\begin{figure}[h]
\includegraphics[scale=0.35]{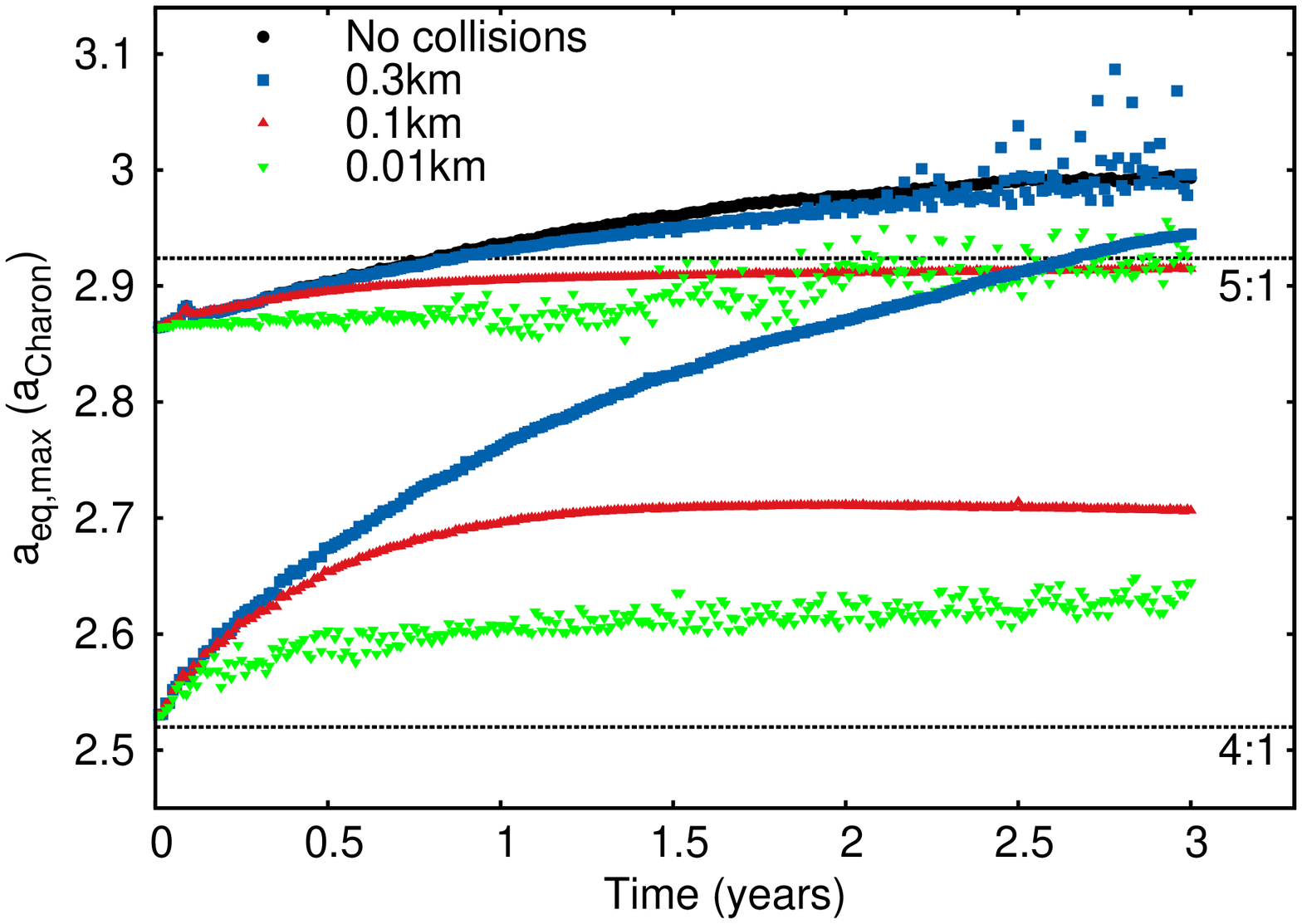} 
\caption[]{The top set of lines are the same as in Figure
  \ref{5to1SB}. The bottom set of data is the identical experiment
  with similar disk and particle properties, but with the disk
  starting closer to Pluto-Charon and evolving in between resonances,
  rather than starting near the 5:1~MMR.
\label{5to1res}}
\end{figure}

What about the resonances? In Figure \ref{5to1res} we present results
of similar disks of debris starting at a different semimajor axis
relative to the barycenter of the Pluto-Charon system. Here the
systems are evolving between the 4:1 or the 5:1 MMR. The results are
slightly different. These disks start closer to Charon, and the
increased perturbations are evident with more extreme movement of the
system's angular momentum for a given particle size. However, the
$R=0.1$~km case moves more than the $R=0.01$~km case, as would be
expected given the two order of magnitude increase in collision
frequency in the latter case. For the test cases further from Charon
(starting between 2.8--2.9 $a_\mathrm{eq}$), the $R=0.1$~km case and
the $R=0.01$~km case evolve to the same spot -- near the
5:1~MMR. While the evolution of the $R=0.3$~km case rapidly moved past
the resonance, both of the more collisionally active cases show signs
of resonant interactions with Charon. The smaller radii case had much
more eccentricity damping and therefore a decreased semimajor axes. As
the semimajor axes decreased the particles were converging with Charon
allowing for resonant interactions (see Cheng 2011).

[EDITOR: Place Fig \ref{5to1res} here]

A similar series of test cases were created to provide a simplistic
representation of a disrupted satellite (dubbed {``simple
  disrupted satellite''}). Here debris shares a similar point of
origin on the orbit but has a range of $a$ and $e$ due to different
initial ejection velocities away from the disrupted bodies orbit.
Particles were distributed on orbits with a range of
$a$ = 2.6--4.0~$a_\mathrm{Charon}$ and $e$ = 0.5--0.65, but with similar
pericenter values ($q \sim 1.35~a_\mathrm{Charon}$) and similar
longitude of pericenter value.  In this test {\it all of the particles
  were initially on unstable orbits}, therefore without collisional
interactions they would all be ejected from the system on very short
timescales ($<10$~years) despite Charon being on a zero eccentricity
orbit. The particles had similar collisional properties described
above ($R=0.1$~km, with no growth or fragmentation).

The behavior of this disk is similar to the first test case, with the
system rapidly ($<3$~years) damping into a ring-like structure, this
time near the 4:1~MMR. This system also experienced a large increase
in angular momentum, circularizing near $\sim$2.5~$a_\mathrm{Charon}$,
a substantial increase from the initial value of
$\sim$2.25~$a_\mathrm{Charon}$.

\begin{figure}[!ht]
\centering
\begin{minipage}[r]{0.95\linewidth}
\centering
\includegraphics[scale=0.35]{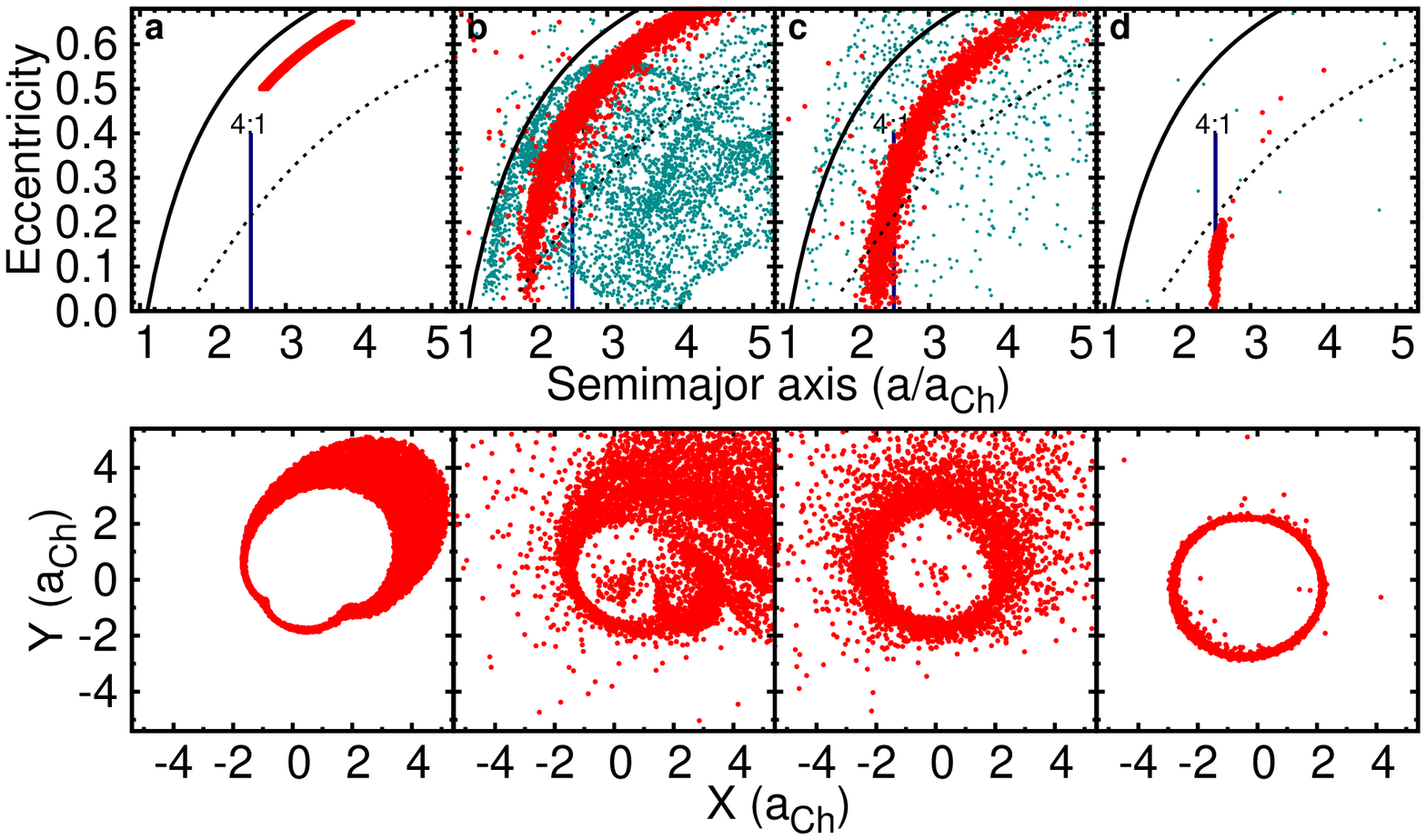} 
\end{minipage}
\hspace{0.01truein}
\begin{minipage}[r]{0.95\linewidth}
\centering
\renewcommand{\baselinestretch}{0.9}
\caption[]{Frames showing the collisional damping of a disk of debris
  with initially correlated periapsis $q$, and a range of $a$ and $e$
  at times (from left to right) 0, 0.1, 0.5, and 3.0~yr. All of the
  particles are initially on unstable orbits and with semimajor axes
  exterior to the 4:1~MMR. The angular momentum of the system
  increases due to kicks from Charon and also the resonant
  interactions of the ring with Charon's orbit. Most particles cross
  the 4:1~MMR while damping semimajor axis, and the final ring
  structure is resonanting with Charon in the 4:1 MMR. The cyan
  particles (smaller point sizes in the top frames only) are from a
  simulation with the same initial conditions, but with no
  collisions. They are rapidly dispersed and lost from the system.
\label{MVB}}
\end{minipage}
\end{figure}

[EDITOR: Place Fig \ref{MVB} here]

As before, this same disk of debris was also modeled without
collisions. The resulting evolution resulted in the ejection or
accretion by Charon of over 60\% of all the particles over the course
of the simulation and no coherent ring-like structure (see small,
cyan, particles in upper panes of Fig \ref{MVB}).

These two simplistic test cases explored the powerful effects of
collisional damping on the dynamical evolution of an eccentric ring.
The tests showed that combined affects of dynamic perturbations from
Charon and collisional damping of an eccentric ring can lead to
outward movement of material in the system. This was one of the main
problems in understanding the suite of small satellites in the
Pluto-Charon system and this is a viable solution. While these
  tests only considered one value of disk semimajor axes and only one
  orbit for Charon (with different eccentricities in each test), the
  outcome of outward movement of material would scale to an epoch
  where Charon was closer to Pluto and still tidally evolving
  outward. In the case where a satellite was disrupted during the
  tidal evolution this outward movement would be very useful in
  helping to push the satellite outward without relying on any
  resonances to move material long distances in the system.

However, these simulations are missing the important physics of
fragmentation and accretion that are necessary to understand where the
disrupted satellite will re-build after it damps and moves out. High velocity
collisions could create swarms of small debris that would dramatically
change the collisional damping timescales of the large particles,
while accretion during low relative velocity collisions could grow a
few or many large bodies from the entire system. Where these bodies
grow will show whether this mechanism can answer the second open
question about the system, as to why the small satellites are located
near resonances.

\section{Disruption of a Primoridial Satellite - Including Fragmentation and Growth}\label{s:lipad}

As shown in Section \ref{s:coldisk}, collisional interaction between
debris orbiting Pluto/Charon can radically change the collective
dynamical behavior of an eccentric ring. Here we test similar
scenarios with growth and accretion aiming to see if the collisionally
evolving rings will preferentially grow near resonances, and we
  expand the study to investigate a wider range of initial ring
  locations relative to the orbit of Charon. There are a few sources
of such debris that could have played a role in the history of the
Pluto-Charon system.  Specifically, the simulations of Canup
(2005,2011) find significant amounts of debris orbiting the
Pluto-Charon system following their giant impact.  Any debris that
avoids accretion by Charon, survives dynamical ejection and accretes
into satellites are then at risk of later dynamical ejection from the
system during the outward tidal evolution of Charon (Lithwick and Wu
2008, Cheng 2011). The mechanism explored in the preceeding
Section can help to move material outward and avoid ejection but
depends on the disruptions of satellites on similar timescales to the
tidal evolution.

Mutual collisions between satellites are a possible means to disrupt
bodies.  Any satellites could have their eccentricities excited by
entering or crossing MMRs, leading to crossing orbits and collisions
(Cheng 2011, Cheng et al. 2014b). Collision velocities would be on order of a few
10's~m/s and approaching 100's m/s as eccentricities get very high
(Nix's orbital velocity is $\sim$140~m/s). However, lower velocity
collisions may simply lead to accretion rather than disruption (a
$R=50$~km target needs to be hit by a $R\sim 43$~km projectile at
100~m/s to disrupt, using the $\bar{Q}^{\star}_\mathrm{D}$ calculation
from Benz and Asphaug 2009).

Heliocentric impactors are another method for disrupting any existing
satellites and producing a significant amount of debris.  In the
environment where the collision probability for Charon's formation was
unity ({\it and nearly every large KBO appears to have suffered
  similar impacts}), then the lifetime of the smaller satellites
($R\sim10-100$~km) could be much shorter than the time to dynamically
deplete the Kuiper Belt (Levison et al. 2008) and possibly shorter
than Charon's tidal evolution timescale. The distribution of relative
velocities for a heliocentric impactor depends on the dynamical state
of the Kuiper Belt and could range from $\sim$100--500~m/s (Pires dos
Santos et al. 2012).

We start this series of calculations by generating the debris from the
disruption of a satellite impacted by an object from an heliocentric
orbit for use as initial conditions for our LIPAD simulations.  We
relied on an approximation by way of a $N$-body simulation of a 39~km
radius object striking a 89~km radius body at 500~m/s with an impact
angle of approximately 45~deg. The target body was made of 9965
discrete, spherical and unbreakable particles and the impactor of 717
particles. The disruption event was modeled with the gravitational and
granular mechanics code {\tt pkdgrav}, which is commonly used for
low-velocity impact modeling (see Leinhardt et al. 2000 and Richardson
et al. 2000). This collision was done in a frame centered on the
target-impactor center of mass. It was then translated into a frame
relative to that of a satellite on different orbits around
Pluto/Charon and with different orientations of the impact
direction. This allowed for the exploration of collisions with a wide
range of geometry relative to the velocity vector of the
satellite. The specific collision modeled here is not derived from a
collisional model, rather it was designed to be very general and have
a high enough resolution and violent enough disruption to model the
relevant physics in a wide range of cases.

\begin{figure}[h]
\includegraphics[angle=0,width=3.5in]{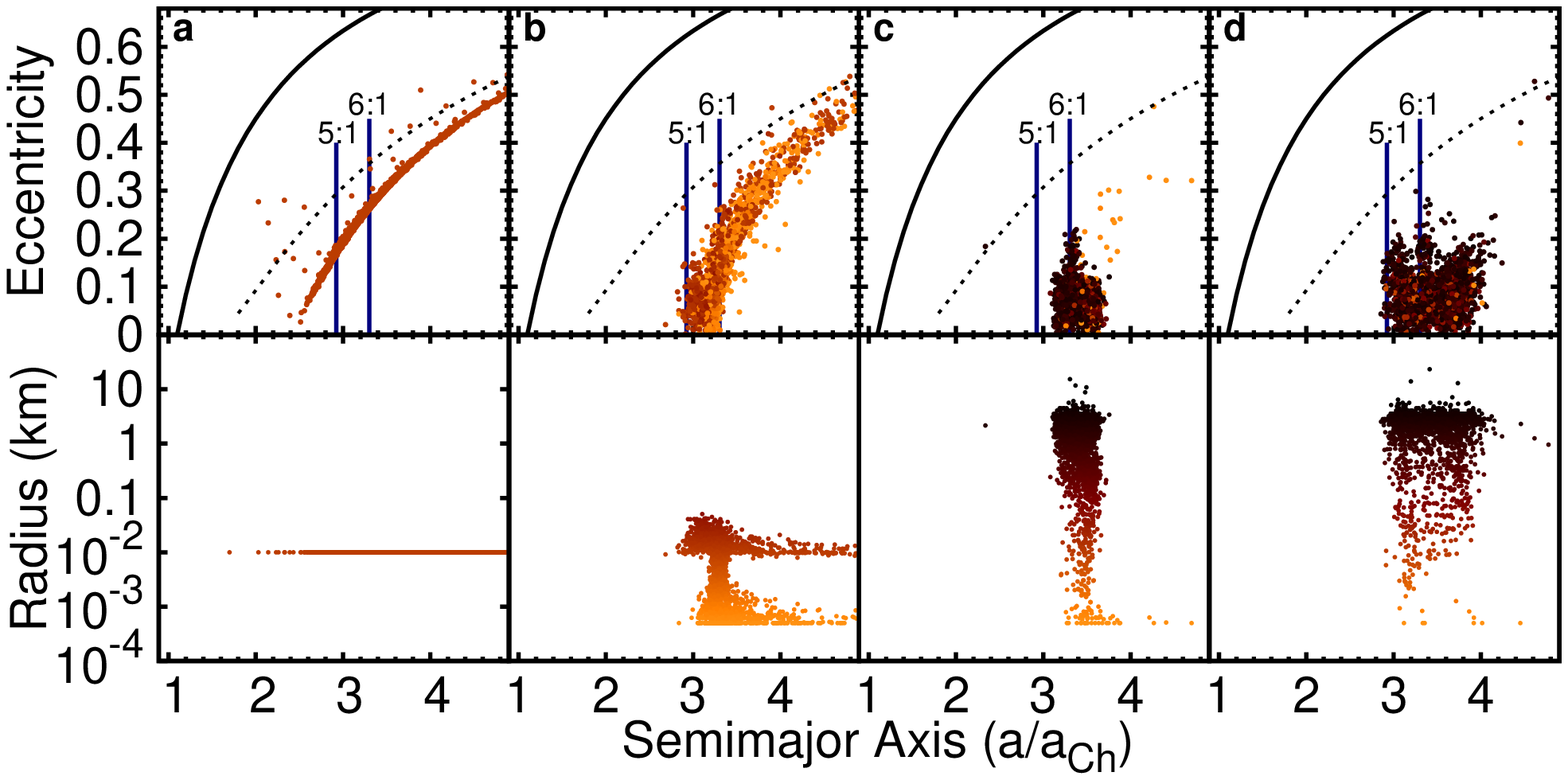}
\caption{Evolution of a disrupted satellite, where Charon has an
  eccentricity of 0.2, and the satellite was disrupted at 48,000~km
  from the Pluto-Charon barycenter on a zero eccentricity orbit. The
  dotted line in the top panels signifies the stability boundary, the
  solid line marks where an orbit's pericenter crosses Charon and the
  major MMRs are labeled. The bottom panels shows the size of the
  particles, where here they were initially 0.01~km, and were allowed
  to break to a minimum size of 10$^{-4}$~km. The colors represent
  their radius, with particles above 1~km being black. The plots
  represent the evolution of the system at 0.0, 5, 100, and 400
  years.
\label{figHM5_48_e2}}
\end{figure}

The nature of the debris field causes a thin ``tail'' of debris that
generally share a very similar pericenter ($q$) at the breakup
location with a range of distributed, but correlated, $a$ and $e$
(similar configuration to the second of our simple disrupted satellite
tests demonstrated above; Fig. \ref{figHM5_48_e2}a shows the collision
outcome translated to the frame orbiting Pluto-Charon). The geometry
tested most frequently in this work was when the impactor's velocity
vector was aligned with the target's (satellite) velocity vector
around Pluto-Charon. With this geometry some of the mass was
immediately on orbits escaping the Pluto-Charon system, but most of
the mass was on moderate eccentricity orbits ($e<0.4$) within
semimajor axis $\sim$4~$a_\mathrm{Charon}$. In the opposite geometry,
when the impact velocity was anti-aligned with the satellite velocity,
most mass was rapidly ejected from the system due to decreased
pericenter distances and subsequent close encounters with Pluto or
Charon.

In our main simulations we employed the complete fragmentation and
growth capabilites of {\tt LIPAD}. The simulations were started with
4096 {\it tracer} particles, each of mass 1.86$\times$10$^{14}$~kg,
totaling 7.6$\times10^{17}$~kg in the system (Tholen et al. 2008
estimated a mass for Nix of 6$\times10^{17}$~kg). The smallest radius
that a fragment could attain during a collision was
5$\times10^{-4}$~km, and the collisional routines used the Benz and
Asphaug (1999) $Q^{\star}$ law for ice (line 4 in their
Table~III). The simulation used timesteps of 4$\times10^{-5}$~years,
or about 21~minutes, which is $\sim$437 timesteps per orbit of Charon.

This formula of taking the debris from a collision and placing them on
a satellite's orbit was repeated for different eccentricities of
Charon, 0.0--0.3, and for different satellite semimajor axes
$a~=$~2.2--2.8~$a_\mathrm{Ch}$. This grid of simulations was designed
to test for the preferential growth of satellites in or near MMRs for
a wide range of Charon orbital properties. Computational limitations
did not allow for simulations to extend for the $\gg$1000 years that
might be necessary to allow each simulation to evolve to a single or
stable system of satellites. Instead we consider the location of the
angular momentum of the system of debris at the end of each run
(typically at 200~years), as each run typically collisionally damped
to a ring on the order of a few years allowed for the growth of large
bodies in tens of years.

In all, 24 simulations were run to cover this parameter space, with
significantly more test runs to examine the sensitvity of each to the
various simulation parameters. The typical behavior is shown in Figure
\ref{figHM5_48_e2}, where the particles inititally have a radius of
0.01~km (some tests were run over an order of magnitude range of
initial sizes and also using an initial size distribution with minimal
differences in outcomes). The first few years were dominated by
fragmentation where many particles grind down to the minimum allowable
size. This is followed by a period of accretion due to their increased
collisional damping leading to lower impact velocities (see the column
forming in Fig \ref{figHM5_48_e2}b).

[EDITOR: Fig \ref{figHM5_48_e2} here]

The location of the first collisions resulting in growth was important
for the simulation outcomes (Fig \ref{figHM5_48_e2}c). The growth
always started with the smallest particles as they have damped to very
low eccentricity and experience low velocity collisions with each
other. After the first accretion events there is rapid growth at the
same location, building a ``tower'' structure in radius vs. semimajor
axis space. The growth occurs in a very limited space, the ``tower'',
from this point forward where eventually the largest bodies in the
system are built and most of the mass is in this structure. These
structures are essentially ring structures where the fragmentation and
collisional damping have limited much of the simulation mass to a
narrow range of semimajor axis and subsequent growth at this spot is
inevitable and then quite fast.

In Figure \ref{figHM5_48_e2}c the growth has reached $\sim$10~km sized
particles and most of the mass of the system resides in the larger
($>$1~km) particles with only a tail of smaller debris. The high
eccentricity tail of debris is largely gone, and nearly all material
is on low eccentricity ($e<0.2$) orbits in a narrow range of semimajor
axis. It has damped to a narrow ring. There are a few particles that
appear to hug the 6:1~MMR at slightly higher eccentricity, suggesting
that they are possibly being excited due to resonant interactions.

Finally in Figure \ref{figHM5_48_e2}d the ring has built enough
large particles that it is now diffusing in semimajor axis. There are
still a few particles seemingly excited by the 6:1~MMR, and also some
particles that appear to have diffused inward and started interacting
with the 5:1~MMR. The state of this simulation shows the complication
in determining the endstate of these simulations. The angular momentum
of this system is very near the 6:1~MMR, but the disk is clearly
diffusing and not simply accreting into one or a few satellites. The
build up at the 5:1~MMR may be an important process, but computational
limits have frustrated further investigation.

\begin{figure}[h]
\includegraphics[angle=0,width=3.5in]{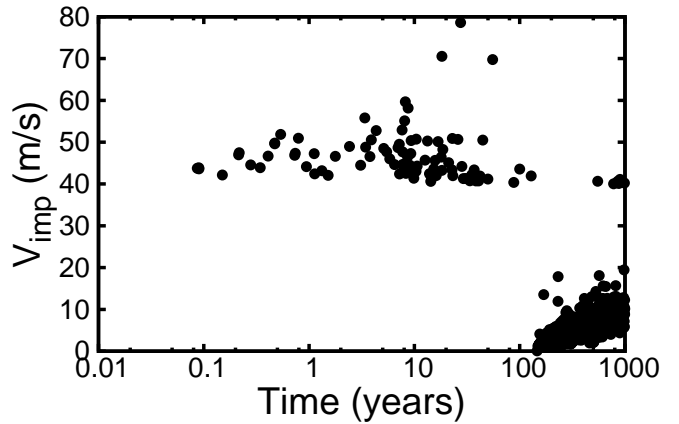}
\caption{Impact velocities as a function of time for a single particle
  in a simulation similar to that shown in Fig \ref{figHM5_48_e2}.
\label{LIPAD_Colls}}
\end{figure}

The collisional environment that produces the evolution in Fig
\ref{figHM5_48_e2} produces fragmentation early when eccentricities
are high, and accretion later after the disk has dynamically cooled
into a ring. In a similar simulation the collisions suffered by one
object were tracked, and the impact velocities as a function of time
are plotted in Fig \ref{LIPAD_Colls}. During the first $\sim$10 years
of the disk's evolution collisions are typically at or above 40~m/s,
which is correlated with the large amounts of fragmentation found in
Fig \ref{figHM5_48_e2}b, where the production of very small fragments
is found. At later times the impact velocities become very low, less
than $\sim$ 10~m/s, as the system has dynamically damped into a ring
with very low eccentricities. From this point, in Fig
\ref{figHM5_48_e2}c accretion is found and correlates with these much
lower velocities. Finally as the simulation approaches 1000 years the
impact velocities slowly increase above 10~m/s, correlated with the
ring spreading into a disk due to the presence of larger bodies
ability to scatter smaller ones with some particles reaching 
higher eccentricities.

[EDITOR: Fig \ref{LIPAD_Colls} here]

\begin{figure}[h]
\includegraphics[angle=0,width=3.5in]{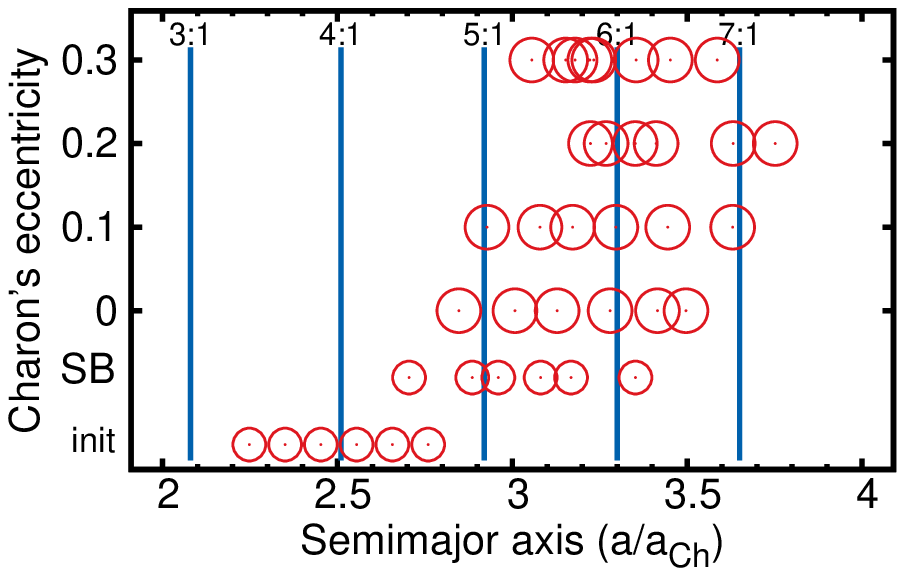}
\caption{Summary of the location of the
  circularized angular momentum of the remaining debris after 200
  years for a range of Charon eccentricity and initial satellite
  location. The y-axis nominally indicates the eccentricity of Charon
  for each represented simulation, with ``SB'' indicating simulations
  where the mass of Charon was added to Pluto and the simulation was
  run around a single massive body. The locations marked by ``init''
  are the locations of the initial disrupted satellite. 
\label{Summary}}
\end{figure}

While some simulations result in rings of material near MMRs, it does
not appear to be a systematic outcome when the final location of the ring's
angular momentum are plotted (see Fig. \ref{Summary}). A set of
control cases were also run where the disruption debris was placed in
orbit around a single central body that had the combined mass of Pluto
and Charon. As expected, this case shows no preference for growth at
any specific locations, and highlights some of the randomness in the
location of the final ring of debris. The spacing and distribution of
the final rings of debris is similar to that produced for both the
cases of Charon eccentricity of 0.0 and 0.1, neither of which show any
preference for growth at a MMR.

The case for Charon eccentricity of 0.3 was expanded and 3 additional
initial conditions were tested, with initial orbits at
$a=$1.9--2.1~$a_\mathrm{Ch}$. As expected, these closer cases were
dynamically kicked outward and build rings at much more distant orbits
 (between the 5:1 and 6:1). The closer initial orbits led to
increased loss of mass from the system as more of the initial ring
distribution was on unstable orbits.

While there is no preference for growth near MMR, the movement outward
of material is clearly seen here, as it was in the simple
collision-only cases demonstrated previously. The tail of debris
extends into regions of $a-e$ space that are unstable and thus there
are substantial perturbations from Charon that drive the whole system
outward. Meanwhile the collisions between disk particles are energetic
enough to fragment particles which leads to substantial populations of
small particles and enhances collisional damping of the system. This
drives each system to damp on orbits further than the initial location
of the disrupted satellite and a net movement outward of the system.

[EDITOR: Fig \ref{Summary} here]

While this clearly shows the outward movement of material in the
  system for all Charon eccentricities examined, including
  $e_\mathrm{Ch}=0.0$, the suite of simulations were only run for one
  specific semimajor axis of Charon, that of today's separation. Given
  that the stability region in $a-e$ space will scale with Charon's
  orbit, the dynamical lifetimes should also scale with Charon's
  orbit. The collisional timescales may change somewhat for closer
  orbits of Charon as the disrupted satellite at the same distance in
  terms of Charon's orbit $a_\mathrm{Ch}$ will fill less volume and
  have an increased collision rate. Our earlier tests explored 3 order
  of magnitude of collision rates and all found outward movement of
  material, so we expect that these results will apply throughout the
  tidal evolution of Charon.

\section{Conclusions and Discussion}\label{s:conc}

This work reports on a series of numerical experiments designed to
understand the origin and evolution of the small satellites of
Pluto. This work found that: 

\begin{enumerate}
\item{}There are regions of stability in $a-e$ space, outside of which
  particles have very short survival timescales in the Pluto-Charon
  system if only gravitational interactions are considered. Collisions
  between particles can stabilize particles initially in the unstable region.

\item{} The combination of perturbations by Charon or particles
  reaching resonant orbits combine to increase angular momentum of a
  collisionally evolving disk of debris, moving the entire system
  outward.

\item{} Satellite disruption, and the subsequent collisional damping and
re-accretion, does not lead to preferential formation in MMRs in the
range of parameters tested in this work.
\end{enumerate}

The satellite system of Pluto remains mysterious.  As stated earlier,
there are two major problems with the small satellites of Pluto, and
we can report progress on one of the two. The first problem is that
the satellites today are much more distant than can be explained by
Charon-formation impact models. Here, by including collisional
evolution in dynamical models, we have found that debris can
experience substantial kicks from Charon while on unstable orbits, but
then return to stable orbits due to collisions with other orbiting
debris.  This effect can result in the movement of material outward in
the system.

The likelihood of this being an important process during the history
of the Pluto-Charon system is not calculated, nor is it a trivial
calculation as it requires a detailed understanding of the excitation,
depletion and collisional evolution of the Kuiper Belt. The tidal
evolution timescale of Charon is on the order of a few million years
and so collisional lifetimes must have been shorter than this for this
to be important. However, it is expected that the Charon-forming
collision happened in a different collision environment than found
today in the Kuiper Belt and collisional timescales for
$\sim$100~km bodies must have been very short. Understanding the
collisional history of both bodies by investigations with the {\it New
  Horizons} spacecraft mission may help to better understand these
issues. This, in effect, suggests that today's small satellites
  are essentially the last in multiple generations of previous
  satellites.

The second problem is the curious configuration of the small
satellites, each near a MMR with Charon. There were indications in
some simulations that collisionally active disks of debris would damp
into a ring sturcture and could be caught while crossing a MMR (as
found fortuitously in the simulation shown in Figure
\ref{figHM5_48_e2}). However, a larger parameter space of simulations
using a full fragmentation and accretion model failed to show a strong
preference at any eccentricity of Charon. A large number of the
simulation parameters were varied with none clearly indicating
importance in a MMR capture mechanism. However, one curious outcome
from these studies was the inner edge of a diffusing ring that
interacts with a MMR. The importance of this effect could not be
investigated here due to computational limitations, but could be
potentially be responsible for building small satellites one MMR
inward of a larger satellite.

Can success be claimed in the first problem and not the second?  If
collisional damping of an eccemtric ring is necessary to move material
outwards should it not also explain the orbital configuration?  It is
certainly possible that collisional evolution as tested here is not
important or that there was another more dominant mechanism that could
both move the satellites or their building blocks and result in their
organized accretion near resonance. It is also possible that we have
uncovered the means to move material outward in the system, and that
there are more or different effects that will ultimately be
responsible for the final orbital configuration.

Is it possible that we are simply being fooled by the system and that
the orbital configuration is just luck? While four satellites near
resonance is hard to explain, both Kerberos and Styx both reside in
relatively narrow regions of orbital stability (Pires dos Santos et
al. 2012, Youdin et al. 2013). Nix and Hydra themselves are near a
resonant configuration with each other, and so it is conceivable that
Kereberos and Styx simply formed in the only places they could in a
system with two more massive satellites that were somehow pushed into
or near a resonance with each other.

Another question that this work can address with the tools
  developed here relates to the very first step of this entire process
  --- the survival of any debris immediately following the formation
  of Charon. In the preceeding Sections, we assumed that the
  post-impact debris measured by Canup (2011) would survive and remain
  in the Pluto-Charon system, despite the typically close orbits
  ($\sim$10~$R_\mathrm{Pluto}$) and the potentially high eccentricity of
  Charon. Using the same simulation configuration described above we
  have done a series of tests to estimate the collision rates
  necessary to stabilize debris at the formation distances found in
  Canup (2011). For a Charon eccentricity of 0.3, we considered disk
  masses of $6.75\times10^{17}, 3.2\times10^{17}$ and
  $1.2\times10^{17}$~kg (Brozovi\'{c} et al. 2014 estimated
  $M_\mathrm{Nix}+ M_\mathrm{Hyd}=9.2\times10^{16}$~kg), placed on
  orbits at approximately $2\times a_\mathrm{Charon}$.  This configuration
  is similar to the $a\sim 10 R_\mathrm{Pluto}$ typically found for
  the debris relative to $a=4-6 R_\mathrm{Pluto}$ for Charon in Canup
  (2011). The simulations used the same collisional debris setup as
  explored in the previous Sections.

Only in the most massive case did significant amounts of debris
survive. For $1.2\times10^{17}$~kg only 5 particles were left at 500
years, and no coherent ring structure ever formed. Increasing to
$3.2\times10^{17}$~kg a very tenuous ring structure formed from the 70
particles that survived for 500 years, but for $6.75\times10^{17}$~kg
case over 800 particles remained, formed a ring, and experienced
significant growth. The largest particle reached 4.8~km, and the
recognizable ``tower'' structure grew between the 6:1 and 7:1~MMR.

These tests spanned the critical regime where the collision rate
became high enough to keep debris in the system.  The collision rates
of 0.006, 0.039 and 0.102 collisions per particle per Charon orbit
were found for the lowest to highest mass cases respectively, and thus
a rate between the latter two values can be considered the critical
limit for survival of debris in this scenario.  The Canup (2011)
simulations found total masses of debris ranging from $10^{17} -
10^{21}$~kg (only 3 of 19 simulations were below $10^{18}$~kg). While
the masses are typically above what was used in this test (with many
simulations with 2--3 orders of magnitude higher mass), the collision
rate will be the important quantity and will depend on the size
distribution of debris in the system.

Looking beyond the Pluto-Charon system, some of the dynamical
  interactions between a massive perturber and a disk or ring of
  debris could be relevant on planetary scales. In our own Solar
  System the scattered disk of Kuiper Belt Objects shows the
  characteristic orbital features of having been excited by Neptune.
  In the context of this work, the dynamically excited scattered disk
  would have collisionally damped if the characteristic collisional
  timescales were shorter than the dynamical lifetimes. 

Beyond our Solar System, there are circum-binary planets
   orbiting stellar systems with mass ratios similar to Pluto-Charon
   (see Kepler-16b reported in Doyle et al. 2011 and Kepler-34b and
   Kepler-35b reported in Welsh et al. 2012). Many of the effects
   driving planet formation in these systems will be different,
   particularly effects of the gaseous stellar nebula (see Meschiari
   2014), but some of the orbital perturbations on the planetary
   building blocks will be of similar magnitude as those on
   satellite building blocks around Pluto and Charon.

In summary, this work has made progress on part of the confounding
problems of the small satellites of Pluto. Hopefully this and other
recent works, when combined with a very detailed study of the system
by way of the {\it New Horizons} spacecraft mission (Stern 2008), will
help to solve some of these outstanding mysteries.

\acknowledgements{ K.J.W and H.F.L acknowledges support from NASA’s
  NLSI (NNA09DB32A) and SSERVI (NNA14AB03A) programs that supported
  code development and H.F.L acknowledge support from NASA's OPR and
  OSS programs. This work used the Extreme Science and Engineering
  Discovery Environment (XSEDE), which is supported by National
  Science Foundation grant number ACI-1053575.}

\end{document}